# Dissolving and Stabilizing Soft $WB_2$ and $MoB_2$ Phases into High-Entropy Borides via Boron-Metals Reactive Sintering to Attain Higher Hardness

Mingde Qin, Joshua Gild, Haoren Wang, Tyler Harrington, Kenneth S. Vecchio, Jian Luo[*]

Department of NanoEngineering; Program of Materials Science and Engineering, University of California, San Diego, La Jolla, CA, 92093, USA

**Abstract:**

Five single-phase $WB_2$- and $MoB_2$-containing high-entropy borides (HEBs) have been made via reactive spark plasma sintering of elemental boron and metals. A large reactive driving force enables the full dissolution of 10-20 mol. % $WB_2$ to form dense, single-phase HEBs, including $(Ti_{0.2}Zr_{0.2}Hf_{0.2}Mo_{0.2}W_{0.2})B_2$, $(Ti_{0.2}Ta_{0.2}Cr_{0.2}Mo_{0.2}W_{0.2})B_2$, $(Zr_{0.2}Hf_{0.2}Nb_{0.2}Ta_{0.2}W_{0.2})B_2$, and $(Zr_{0.225}Hf_{0.225}Ta_{0.225}Mo_{0.225}W_{0.1})B_2$; the successful fabrication of such single-phase $WB_2$-containing HEBs has not been reported before. In the processing science, this result serves perhaps the best example demonstrating that the phase formation in high-entropy ceramics can strongly depend on the kinetic route. A scientifically interesting finding is that HEBs containing softer $WB_2$ and/or $MoB_2$ components are significantly harder than $(Ti_{0.2}Zr_{0.2}Hf_{0.2}Nb_{0.2}Ta_{0.2})B_2$ (with harder binary boride components). This exemplifies that high-entropy ceramics can achieve unexpected properties.

---

[*] Corresponding author. E-mail address: jluo@alum.mit.edu (J. Luo).

# 1. Introduction

In the physical metallurgy community, high-entropy alloys have received great attentions since their introduction by Yeh et al. [1] and Cantor et al. [2] in 2004. In the ceramics community, high-entropy rocksalt [3], perovskite [4], fluorite [5], and pyrochlore [6] oxides, as well as borides [7], carbides [8-11], silicides [12, 13], and fluorides [14] have been made in the past a few years. Recently, it was proposed to further broaden high-entropy ceramics (HECs) to compositionally-complex ceramics (CCCs) [15, 16], where medium-entropy and/or non-equimolar compositions can often outperform their equimolar, high-entropy counterparts.

In 2016, Gild et al. [7] first reported the fabrication of single-phase high-entropy borides (HEBs) in the hexagonal $AlB_2$ structure, which represent the first non-oxide high-entropy ceramics fabricated in bulk form and a new class in ultra-high temperature ceramics (UHTCs). Since then, substantial efforts have been made to fabricate HEBs [17-22]. Notably, Tallarita et al. synthesized $(Ti_{0.2}Zr_{0.2}Hf_{0.2}Nb_{0.2}Ta_{0.2})B_2$ [20] and $(Ti_{0.2}Zr_{0.2}Hf_{0.2}Nb_{0.2}Mo_{0.2})B_2$ [22] via a two-step processing method, achieving relative densities of ~92.5% for both cases; in their method, elemental metals and boron were first converted to metal diborides via self-propagating high-temperature synthesis (SHS), and subsequently densified by spark plasma sintering (SPS); noting that although SPS is the most commonly used terminology in literature, it more accurately described as direct current sintering (DCS) or field assisted sintering technique (FAST). Tallarita et al. also tried directly reactive SPS $(Ti_{0.2}Zr_{0.2}Hf_{0.2}Nb_{0.2}Mo_{0.2})B_2$ from boron and metals in one step (1950 °C, 20 min, 20 MPa), but did not achieve a high density and uniform microstructure due to out-gassing [22]. Moreover, direct reactive SPS from elemental precursors was applied to make high-entropy $(Ti_{0.2}Zr_{0.2}Nb_{0.2}Ta_{0.2}W_{0.2})C$ [23] and $(Ti_{0.2}Zr_{0.2}Nb_{0.2}Mo_{0.2}W_{0.2})Si_2$ [12].

$WB_2$ of the $AlB_2$ prototype structure is thermodynamically unstable and difficult to synthesize [24, 25]. Instead, a $W_2B_{5-x}$ phase (space group $P6_3/mmc$, no.194), a different hexagonal structure of stacking $AlB_2$-type layers separated by defect-rich puckered boron layers of alternating orientations, is the equilibrium phase in the W-B phase diagram. Only a limited number of studies demonstrated the formation of $AlB_2$-type $WB_2$ in nano-scale coating or thin films [26-29] or after extensive ball milling [30]. Notably, an $AlB_2$-type phase can be kinetically stabilized in $TiB_2$-$WB_2$ [31, 32]; yet, a WB secondary phase found to precipitate out in $TiB_2$-$CrB_2$-$WB_2$ [33]. In recent studies of HEBs, a W-rich secondary phase was reported to form in nominally $(Ti_{0.2}Zr_{0.2}Hf_{0.2}Mo_{0.2}W_{0.2})B_2$ specimens made by either SPS of a high energy ball milled (HEBM) mixture of five binary borides [7] or powders synthesized by boro/carbothermal reduction [21]. To the best of our knowledge, single-phase HEBs containing 10-20 mol. % $WB_2$ have not been successfully made to date (albeit HEBs containing 20 mol. % $MoB_2$ have been made [7, 18, 22], although $AlB_2$-type $MoB_2$ is also not a stable phase at room temperature).

In this study, we showed that reactive SPS of elemental metal and boron powders enables the synthesis of three compositions of single-phase HEBs containing 20 mol. % $WB_2$ (plus another HEB containing 10 mol. % $WB_2$); in contrast, SPS of milled mixtures of five binary borides always led to the formation of a secondary WB-rich monoboride phase in the same four compositions. Notably, we further made a scientifically interesting discovery: HEBs containing softer $WB_2$ and/or $MoB_2$ components [34] are harder than $(Ti_{0.2}Zr_{0.2}Hf_{0.2}Nb_{0.2}Ta_{0.2})B_2$. Thus, this study exemplifies that HECs can achieve unexpected properties beyond the simple mixture effect.

## 2. Method

For boron-metals reactive SPS, elemental powders of Ti, Cr, Zr, Nb, Mo, Hf, Ta, and W (>99% purity, ~325 mesh, purchased from Alfa Aesar, MA, USA) and boron (99% purity, 1-2 μm, purchased from US Research Nanomaterials, TX, USA) were utilized for making specimens HEB-A1 to HEB-F1 listed in Table 1. For each composition, appropriate amounts of metals and boron were weighted out in batches of 5 g; noting that we added 10% excess B (i.e., weighted a nominal metal-to-boron atomic ratio of 1:2.2) to offset the loss due to reaction with native oxide and subsequently evaporation. The powders were hand-mixed and subsequently HEBM in a Spex 8000D mill (SpexCertPrep, NJ, USA) in tungsten carbide lined stainless steel jars and 10 mm tungsten carbide milling media (ball-to-powder ratio ≈ 4:1), for 50 min with 1 wt. % (0.05 g) stearic acid as lubricant. The HEBM was performed in an argon atmosphere ($O_2$ < 10 ppm) to prevent oxidation.

The milled powders were loaded into 10 mm graphite dies lined with graphite foils in batches of 2 g, and subsequently consolidated into dense pellets via SPS in vacuum ($10^{-2}$ Torr) using a Thermal Technologies 3000 series SPS (CA, USA). During the initial temperature ramping at 50 °C/min with uniaxial load of 10 MPa, in-situ reaction of metals and boron likely takes place. Differing from Tallarita et al.'s trial of directly reactive SPS of $(Ti_{0.2}Zr_{0.2}Hf_{0.2}Nb_{0.2}Mo_{0.2})B_2$ in one step (1950 °C, 20 min, 20 MPa) that did not achieve a high density [22], our specimen was first held in SPS at 1400 °C and 1600 °C, successively, for 40 min each, to allow out-gassing and reduction of native oxides with excess boron [35, 36]. After that, the temperature was raised to 2000 °C at a lower heating rate of 30 °C/min and sintered isothermally for 10 min for final densification; at the same time, the load was increased to 50 MPa at a rate of 5 MPa/min. Finally, the sintered specimen was cooled in the SPS machine to room temperature.

For a comparison study, specimens of the four W-containing compositions were also synthesized via conventional HEBM and SPS of five binary metal borides (following the route in Ref. [7]), and these benchmark specimens are labelled as HEB-C2 to HEB-F2. $TiB_2$, $CrB_2$, $ZrB_2$, $NbB_2$, $HfB_2$, $W_2B_{5-x}$ (>99% purity, ~325 mesh, purchased from Alfa Aesar, MA, USA), $Mo_2B_{5-x}$, and $TaB_2$ (>99% purity, 45 μm, purchased from Goodfellow, PA, USA) were utilized as the precursors for these specimens; and it should be noted that $Mo_2B_{5-x}$ and $W_2B_{5-x}$ powders were used as the starting binary borides of Mo and W since the $AlB_2$-structured $MoB_2$ and $WB_2$ are unstable and not commercially available.

All sintered specimens were ground (to remove carbon contamination from graphite tooling) and polished before further characterization. Densities for specimen HEB-A1 to HEB-F1 were measured via Archimedes' method; their relative densities were calculated based on theoretical densities calculated from the ideal stoichiometry and the lattice parameters measured by X-ray diffraction (XRD) and listed in Suppl. Table S1. XRD characterizations were carried out on all specimens using a Rigaku Miniflex diffractometer with Cu Kα radiation at 30 kV and 15 mA. Scanning electron microscopy (SEM), electron dispersive X-ray spectroscopy (EDS), and electron backscatter diffraction (EBSD) data for were collected with a FEI Apreo microscope equipped with an Oxford N-Max[N] EDX detector and an Oxford Symmetry EBSD detector. Vickers' microhardness tests were carried out on a diamond indenter with loading force of 1.96 N (200 gf) and holding time of 15 seconds according to ASTM C1327. Over 50 measurements were conducted at different locations to ensure statistical validity and minimize the microstructural and grain boundary effects.

## 3. Results and Discussion

XRD showed that all six specimens HEB-A1 to HEB-F1 synthesized via boron-metals reactive SPS have virtually single $AlB_2$-structured HEB phases without any detectable secondary phase (Fig. 1(a)). Additional XRD patterns of the as-milled metal and boron powders are shown in Suppl. Fig. S2, which shows that the HEB phase formed during the reactive SPS. SEM further showed that these specimens are dense with <3 vol. % of total porosity and/or extra boron (Fig. 2), which is consistent with measured relative densities (95.5% for HEB-D1 and >97% for all other five specimens as shown in Table 1). EDS elemental maps shown in Fig. 2(a-1) to (f-1) confirm the formation of homogeneous HEB solid solutions for all six compositions.

Thus, the above results showed that reactive SPS of elemental metals and boron enabled the formation of four $WB_2$-containing single-phase HEBs without any detectable secondary phase, including HEB-C1: $(Ti_{0.2}Zr_{0.2}Hf_{0.2}Mo_{0.2}W_{0.2})B_2$, HEB-D1: $(Ti_{0.2}Ta_{0.2}Cr_{0.2}Mo_{0.2}W_{0.2})B_2$, HEB-E1: $(Zr_{0.2}Hf_{0.2}Nb_{0.2}Ta_{0.2}W_{0.2})B_2$, and HEB-F1: $(Zr_{0.225}Hf_{0.225}Ta_{0.225}Mo_{0.225}W_{0.1})B_2$.

In addition, two $WB_2$-free single-phase HEBs, HEB-A1: $(Ti_{0.2}Zr_{0.2}Hf_{0.2}Nb_{0.2}Ta_{0.2})B_2$ and HEB-B1: $(Ti_{0.2}Zr_{0.2}Nb_{0.2}Ta_{0.2}Mo_{0.2})B_2$, have also been fabricated by boron-metals reactive SPS as benchmark. It is known that these two compositions can also be made into single-phase HEBs via conventional HEBM and SPS of five binary metal borides, as shown in a prior study [7].

In a critical comparison, XRD patterns in Fig. 1(b) show that the four $WB_2$-containing specimens HEB-C2 to HEB-F2 synthesized by conventional HEBM and SPS of five binary borides all exhibit a primary $AlB_2$-structured HEB phase with a secondary phase of orthorhombic (CrB prototype, space group Cmcm, no. 63) monoboride. Small amounts of (Zr, Hf)$O_2$ native oxides were also detected in HEB-C2, HEB-E2, and HEB-F2 where $ZrB_2$ and $HfB_2$ were utilized as precursors (while the specimens HEB-C1, HEB-E1, and HEB-F1 of the same compositions made by reactive SPS of elemental boron and metals are essentially oxide-free, as shown in Fig. 1(a)). EDS elemental maps in Fig. 2(c-2) to (f-2) further showed that the secondary monoboride phases are W-rich. Furthermore, the compositions of these monoboride phases were measured by EDS point analyses to be $(W_{0.78}Mo_{0.19}Ti_{0.02}Hf_{0.01})B$ in HEB-C2, $(W_{0.41}Mo_{0.28}Cr_{0.18}Ta_{0.09}Ti_{0.04})B$ in HEB-D2, $(W_{0.54}Ta_{0.15}Hf_{0.13}Nb_{0.11}Zr_{0.07})B$ in HEB-E2, and $(W_{0.55}Mo_{0.36}Ta_{0.09}Hf_{0.01})B$ in HEB-F2, respectively (Suppl. Fig. S4). Noticeably, the IVB elements Ti, Zr and Hf all have low fractions in these monoboride phases, because their corresponding monoborides are not stable [37]. The higher concentrations of VIB elements Cr, Mo, and W than the VB elements Nb and Ta can be justified by the differences in their formation enthalpies between CrB-type and $AlB_2$-type structures. The differences in formation enthalpies ($H_{MB}^M - H_{MB_2}^M$) are -0.277, -0.211, -0.186, -0.163, and -0.07 in eV/atom, respectively for $M$ = W, Mo, Cr, Ta, and Nb, respectively, based on the Material Project Database [38]. This order largely confirms the relative fractions of these metals in the secondary monoboride phase, where the most negative $H_{WB}^W - H_{WB_2}^W$ corresponds to the highest tungsten concentrations in the monoboride phase (followed by Mo, Cr, and Ta).

The combination of XRD, SEM, EDS and density measurements showed that four dense, $WB_2$-

containing, single-phase specimens HEB-C1 to HEB-F1 have been successfully synthesized via this new reactive SPS of elemental boron and metals route, which could not be fabricated by conventional HEBM and SPS of mixtures of five binary borides (HEB-C2 to HEB-F2). While one may argue that the use of nonstoichiometric $W_2B_{5-x}$ powder might cause a difference here, we note that a W-rich monoboride secondary phase was also observed in a $(Ti_{0.2}Zr_{0.2}Hf_{0.2}Mo_{0.2}W_{0.2})B_2$ (i.e. composition HEB-C) specimen synthesized via boro/carbothermal reduction of mixed oxides and $B_4C$ and subsequent SPS in a most recent study [21] (while the compositions HEB-A and HEB-B were also made into single-phase HEBs via that boro/carbothermal reduction route [21]).

The stabilization of 20 mol. % $WB_2$ into the three compositions (plus 10 mol. % $WB_2$ into another composition) of single-phase HEBs represents an achievement of this work, which have not been achieved in any prior study. This can be explained by the large reactive thermodynamic driving forces with elemental boron and metals as the starting powders. It is yet unknown whether these $WB_2$-containing single-phase HEBs are truly thermodynamic stable, or metastable, phases. Nonetheless, this work serves as excellent demonstration that the phase formation in HEBs can strongly depend on the kinetic (synthesis and processing) route.

EBSD inverse pole figure orientation maps and their grain size distributions are shown in Fig. 3 for all six specimens synthesized via boron-metals reactive SPS. The averaged grain size varies from $8.6 \pm 4.8$ μm (for HEB-B1) to $20.7 \pm 14.8$ μm (for HEB-F1). Notably, all specimens contain some small grains, which is due to the reactive process (as seen in prior studies of reactive sintering of Ti + B [39] and $TiH_2$ + B [40]). Moreover, preferred grain orientation of (001) plane normal to the direction of the applied pressure is observed, as shown in the inverse pole figure in Suppl. Fig. S5. This texture was further confirmed by comparing the experimental XRD pattern with the calculated XRD pattern (Suppl. Fig. S6). To quantify the degree of this texture, Lotgering orientation factor [41] $f$ was calculated to be from ~0.08 (for HEB-E1 with the least texture) to ~0.15 (for HEB-C1 with the most texture), which represent moderate textures. For comparison, a prior study showed a significant texture in ($f \sim 0.6 - 0.7$) in $TiB_2$ made by a similar in-situ reactive SPS from $TiH_2$ [40]. We find that the formation of this texture in our HEBs is related to the applied pressure. To demonstrate this, an HEB-B specimen was fabricated at a reduced initial loading of <1 MPa via the same routine to achieve minimal texture ($f < 0.01$) (Suppl. Fig. S7 and S8).

Vickers microhardness was measured for all six specimens made by reactive SPS of elemental boron and metals and the results are summarized in Table 1. The measured hardness of HEB-A1 (20.9 ± 1.1 GPa) is consistent with those previously-reported dense specimens of the same compositions $(Ti_{0.2}Zr_{0.2}Hf_{0.2}Nb_{0.2}Ta_{0.2})B_2$ synthesized via different routines [18, 21]. Note that single-phase specimens HEB-A1 and HEB-B1 sintered in this study are harder than those less dense (91-93%) specimens first reported by Gild et al., where the differences can be well explained by the porosity as well as oxide inclusions in those earlier specimens [7].

Most interestingly, the five Mo- and W-containing single-phase specimens HEB-B1 to HEB-F1 are harder than the $MoB_2$- and $WB_2$-free reference specimen HEB-A1: $(Ti_{0.2}Zr_{0.2}Hf_{0.2}Nb_{0.2}Ta_{0.2})B_2$ (Vickers hardness 20.9 ± 1.1 GPa). In comparison, the Vickers hardness of the $MoB_2$-containing specimen HEB-B1: $(Ti_{0.2}Zr_{0.2}Nb_{0.2}Ta_{0.2}Mo_{0.2})B_2$ is increased to 24.9 ± 1.1 GPa. This is consistent with two prior studies [7, 21]. More interestingly, four $WB_2$-containing specimens HEB-C1 to HEB-F1 have been made into single HEB phase for the first time. Three of them (HEB-C1, HEB-E1, and HEB-F1) have even greater Vickers hardness of 26-27.5 GPa, while the relatively lower hardness of HEB-D1 (23.7 ± 1.3 GPa) may be explained from its lower relative density (95.5%).

The fact that $WB_2$- and $MoB_2$-containing HEBs are harder is somewhat surprising and highly interesting since prior DFT calculations predicted $WB_2$ (along with $MoB_2$ and $CrB_2$) to have significantly lower hardness (and to be more ductile) than other diborides ($TiB_2$, $ZrB_2$, $HfB_2$, $NbB_2$, and $TaB_2$) [34]. While the predicted hardness values from DFT calculations are always higher than the measured Vickers hardness from experiments, DFT calculations [34] suggested the following order of hardness: $TiB_2 > ZrB_2 > HfB_2 > NbB_2 > TaB_2 > CrB_2 > MoB_2 > WB_2$ (Suppl. Table S2), which should be mostly accurate. Moreover, based on the DFT predictions [34], the hardness values of four $MoB_2$- and $WB_2$-containing single-phase specimens HEB-B1 to HEB-F1 should be lower than that of the HEB-A1, $(Ti_{0.2}Zr_{0.2}Hf_{0.2}Nb_{0.2}Ta_{0.2})B_2$ by approximately 12%, 12%, 33%, 19%, and 21%, respectively (Table 1). In contrast, the measured Vickers hardness values for HEB-B1 to HEB-F1 are in fact higher than that of the HEB-A1 reference specimen by approximately 19%, 24%, 13%, 28%, and 32%, respectively (Table 1).

Similar beneficial effects with tungsten or molybdenum additions were also evident in "low-entropy" solid solutions of metal diborides. In IVB and VB metal diborides, $MoSi_2$ additions were found to enhance the mechanical properties of $ZrB_2$-, $HfB_2$-, and $TaB_2$-based ceramics [42-44]. WB additions are known to improve the oxidation resistance of $ZrB_2$ significantly [45, 46].

It is possible that this enhanced mechanical property is due to a solution effect of introducing different elements that can enhance different mechanical properties (e.g., ductility vs. modulus) [34] as well as higher valence electron concentrations from introducing VIB elements to contribute more *p* electrons to form *p-d* hybridization with boron [47]. Further modeling study is needed to uncover the exact underlying mechanism. Nonetheless, this clearly demonstrates that high-entropy ceramics can exhibit unexpected/improved properties beyond the simple mixture effect.

## 4. Conclusions

In summary, this study presents a novel method to synthesize single-phase $WB_2$- and $MoB_2$-containing HEBs via reactive SPS of elemental boron and metals; this represents the first report that single-phase HEBs that contains 20 mol. % $WB_2$ can be synthesized. These single-phase HEB specimens have achieved high relative densities with virtually no native oxides. In addition, this study demonstrates that the phase formation in HEBs depends on the kinetic route. In a broader context, it further suggests that reactive sintering of elements with large thermodynamic driving forces as a new route to synthesize single-phase HECs to allow alloying of certain elements that are otherwise difficult to dissolve in significant amounts.

Most interestingly, this study further demonstrates that the incorporation of softer $WB_2$ and/or $MoB_2$ components in HEBs can make them harder (although with larger grain size), thereby suggesting that HECs can attain unexpected properties beyond the simple mixture effect.

**Acknowledgement**: This work is supported by an Office of Naval Research MURI program (Grant No. N00014-15-1-2863; Program Managers: Dr. Eric Wuchina and Dr. Kenny Lipkowitz).

**Appendix A. Supplementary Data**

Supplementary Table S1 and S2 and Supplementary Figures S1-S9 related to this article can be found, in the online version, at doi: xxxxxxxx.

**Table 1.** Summary of the six specimens synthesized via reactive sintering of elemental boron and metals. Experimentally measured Vickers hardness values, along with the theoretical rule-of-mixture (RoM) averages calculated from the those of individual metal diborides from DFT calculations (by Zhou *et al.* [34] and listed in Suppl. Table S2), are also listed. However, it is important to note that the theoretical hardness values calculated by this DFT method are always significantly higher than the measured Vickers hardness values from experiments. For example, the hardness of $ZrB_2$ was predicted to be 41.2 GPa by DFT [34], whereas it was only measured to be 21-23 GPa experimentally [42]. Thus, it is not meaningful to directly compare the experimental Vickers hardness with the theoretical RoM averaged hardness values from the DFT calculations. Instead, we calculated the percentage changes for both experimental and DFT values from reference values of the HEB-A1 $(Ti_{0.2}Zr_{0.2}Hf_{0.2}Nb_{0.2}Ta_{0.2})B_2$ specimen. Comparing these relative percentage changes, it is clearly evident that adding $MoB_2$ and $WB_2$ components makes HEBs harder, despite that the DFT calculations predict lower hardness values. See Suppl. Table S1 for additional data of lattice parameters and theoretical densities.

| Specimen | Compositions | Density (g/cm$^3$) (Relative Density) | Experimental Vickers Hardness (GPa) | Theoretical RoM Averaged Hardness from DFT Calculations (GPa) |
|---|---|---|---|---|
| | | | (% Change from HEB-A1) | |
| HEB-A1 | $(Ti_{0.2}Zr_{0.2}Hf_{0.2}Nb_{0.2}Ta_{0.2})B_2$ | 8.13 (98.1%) | 20.9 ± 1.1 | 35.6 |
| HEB-B1 | $(Ti_{0.2}Zr_{0.2}Nb_{0.2}Ta_{0.2}Mo_{0.2})B_2$ | 7.46 (99.2%) | 24.9 ± 1.3 (+19%) | 31.5 (-12%) |
| HEB-C1 | $(Ti_{0.2}Zr_{0.2}Hf_{0.2}Mo_{0.2}W_{0.2})B_2$ | 8.35 (97.5%) | 26.0 ± 1.5 (+24%) | 31.4 (-12%) |
| HEB-D1 | $(Ti_{0.2}Ta_{0.2}Cr_{0.2}Mo_{0.2}W_{0.2})B_2$ | 8.37 (95.5%) | 23.7 ± 1.3 (+13%) | 24.0 (-33%) |
| HEB-E1 | $(Zr_{0.2}Hf_{0.2}Nb_{0.2}Ta_{0.2}W_{0.2})B_2$ | 9.78 (98.1%) | 26.7 ± 1.1 (+28%) | 28.9 (-19%) |
| HEB-F1 | $(Zr_{0.225}Hf_{0.225}Ta_{0.225}Mo_{0.225}W_{0.1})B_2$ | 9.56 (98.8%) | 27.5 ± 1.1 (+32%) | 28.3 (-21%) |

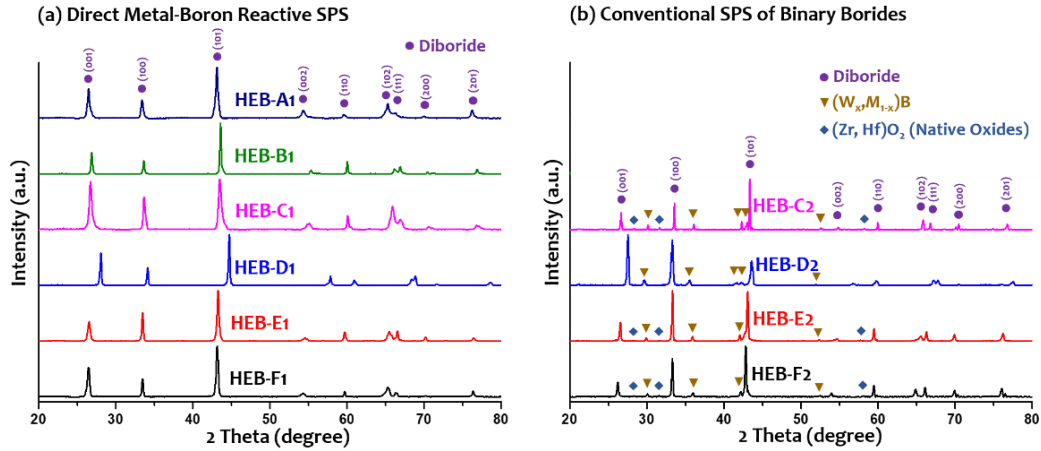

**Fig. 1.** XRD patterns of specimens of different compositions, synthesized via **(a)** reactive SPS of elemental boron and metals and **(b)** conventional SPS of mixtures of binary borides. On the one hand, specimens HEB-A1 to HEB-F1 fabricated via reactive SPS of elemental boron and metals all demonstrate a single AlB$_2$-structured hexagonal phase without detectable secondary phase by XRD. On the other hand, the four W-containing specimens fabricated from mixed binary borides all show noticeable amounts of a WB-rich (as revealed the EDS analysis) secondary monoboride phase, as well as minor (Zr, Hf)O$_2$ native oxides (except HEB-D2). Note that a similar XRD pattern for HEB-C2 was reported in Ref. [7] and included here for comparison purpose.

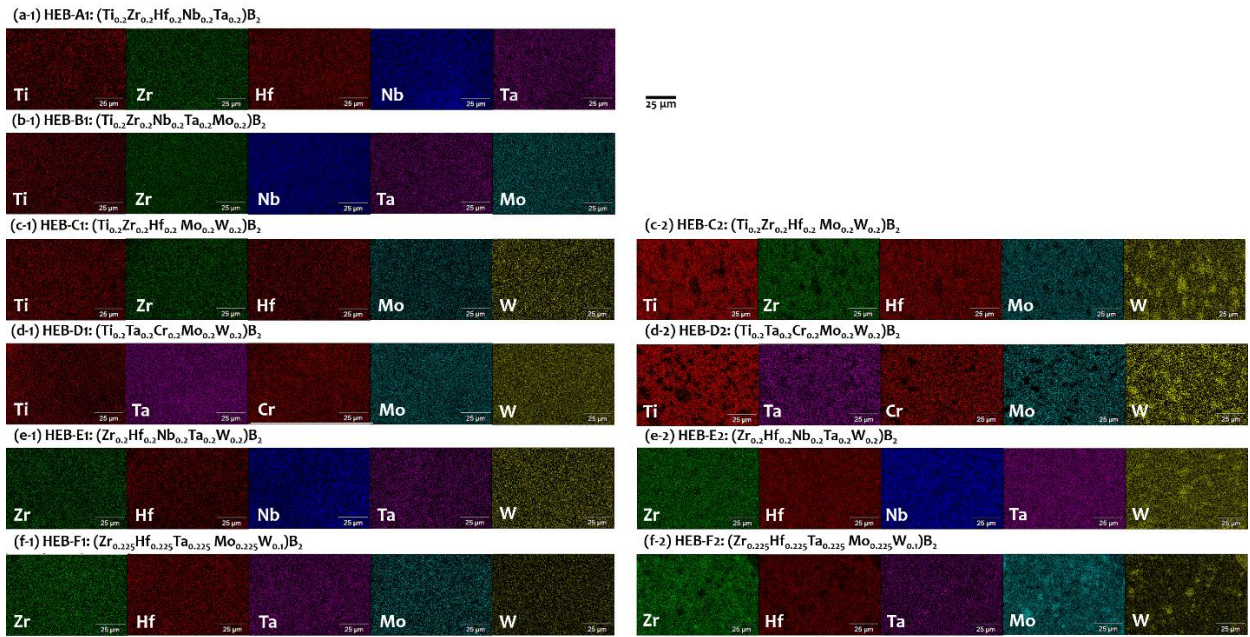

**Fig. 2.** EDS elemental maps of specimens synthesized **(a-1 to f-1)** via reactive SPS of elemental boron and metals and **(c-2 to f-2)** conventional SPS of mixtures of binary borides. On the one hand, all six specimens fabricated by reactive SPS of elemental boron and metals show homogenous elemental distributions. On the other hand, the four W-containing specimens HEB-C2 to HEB-F2 fabricated from SPS of mixtures of binary borides exhibit W-rich secondary phase particles, as well as high porosity Note that similar EDS elemental maps for HEB-C2 was reported in the Suppl. Data for Ref. [7] and included here for comparison purpose. All scale bars represent 25 μm. Enlarged figures of these EDS maps, along with the corresponding SEM micrographs, are shown in Suppl. Fig. S9.

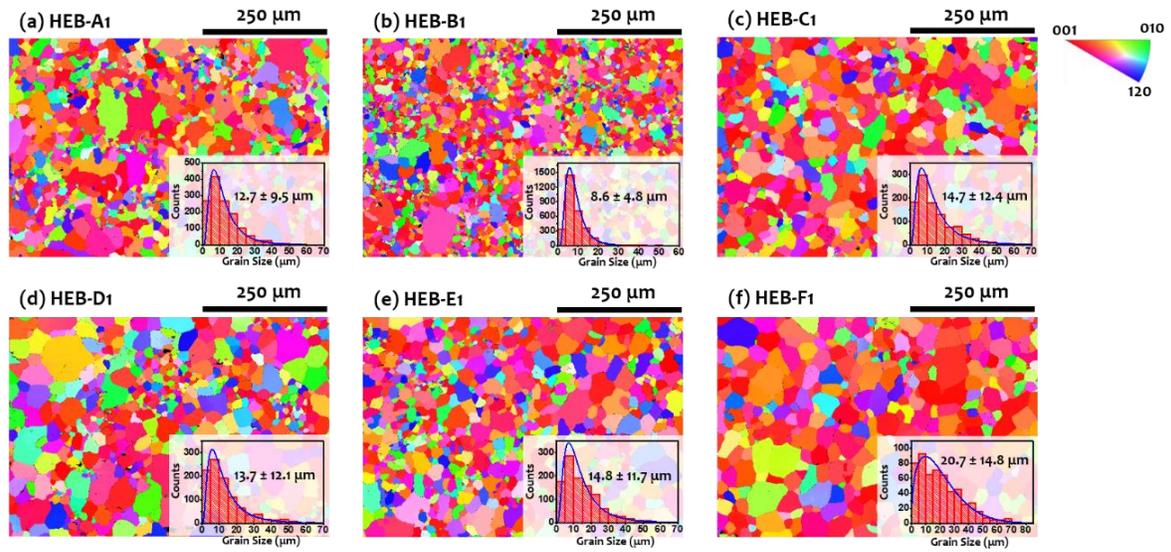

**Fig. 3.** EBSD normal direction inverse pole figure orientation maps for specimens **(a)** HEB-A1, **(b)** HEB-B1, **(c)** HEB-C1, **(d)** HEB-D1, **(e)** HEB-E1, and **(f)** HEB-F1, synthesized via reactive SPS of elemental boron and metals. Insets are grain size distributions.

**References:**

[1] J.-W. Yeh, S.-J. Lin, T.-S. Chin, J.-Y. Gan, S.-K. Chen, T.-T. Shun, C.-H. Tsau, S.-Y. Chou, Formation of simple crystal structures in Cu-Co-Ni-Cr-Al-Fe-Ti-V alloys with multiprincipal metallic elements, Metall. Mater. Trans. A 35(8) (2004) 2533-2536.

[2] B. Cantor, I.T.H. Chang, P. Knight, A.J.B. Vincent, Microstructural development in equiatomic multicomponent alloys, Mater. Sci. Eng. A 375-377 (2004) 213-218.

[3] C.M. Rost, E. Sachet, T. Borman, A. Moballegh, E.C. Dickey, D. Hou, J.L. Jones, S. Curtarolo, J.-P. Maria, Entropy-stabilized oxides, Nat. Commun. 6 (2015) 8485.

[4] S. Jiang, T. Hu, J. Gild, N. Zhou, J. Nie, M. Qin, T. Harrington, K. Vecchio, J. Luo, A new class of high-entropy perovskite oxides, Scripta Mater. 142 (2018) 116-120.

[5] J. Gild, M. Samiee, J.L. Braun, T. Harrington, H. Vega, P.E. Hopkins, K. Vecchio, J. Luo, High-entropy fluorite oxides, J. Eur. Ceram. Soc. 38(10) (2018) 3578-3584.

[6] A.J. Wright, Q. Wang, S.-T. Ko, K.M. Chung, R. Chen, J. Luo, Size disorder as a descriptor for predicting reduced thermal conductivity in medium- and high-entropy pyrochlore oxides, Scripta Mater. 181 (2020) 76-81.

[7] J. Gild, Y. Zhang, T. Harrington, S. Jiang, T. Hu, M.C. Quinn, W.M. Mellor, N. Zhou, K. Vecchio, J. Luo, High-entropy metal diborides: a new class of high-entropy materials and a new type of ultrahigh temperature ceramics, Sci. Rep. 6 (2016) 37946.

[8] P. Sarker, T. Harrington, C. Toher, C. Oses, M. Samiee, J.-P. Maria, D.W. Brenner, K.S. Vecchio, S. Curtarolo, High-entropy high-hardness metal carbides discovered by entropy descriptors, Nat. Commun. 9(1) (2018) 4980.

[9] T.J. Harrington, J. Gild, P. Sarker, C. Toher, C.M. Rost, O.F. Dippo, C. McElfresh, K. Kaufmann, E. Marin, L. Borowski, P.E. Hopkins, J. Luo, S. Curtarolo, D.W. Brenner, K.S. Vecchio, Phase stability and mechanical properties of novel high entropy transition metal carbides, Acta Mater. 166 (2019) 271-280.

[10] X. Yan, L. Constantin, Y. Lu, J.-F. Silvain, M. Nastasi, B. Cui, (Hf0.2Zr0.2Ta0.2Nb0.2Ti0.2)C high-entropy ceramics with low thermal conductivity, J. Am. Ceram. Soc. 101(10) (2018) 4486-4491.

[11] E. Castle, T. Csanádi, S. Grasso, J. Dusza, M. Reece, Processing and properties of high-entropy ultra-high temperature carbides, Sci. Rep. 8(1) (2018) 8609.

[12] Y. Qin, J.-X. Liu, F. Li, X. Wei, H. Wu, G.-J. Zhang, A high entropy silicide by reactive spark plasma sintering, J. Adv. Ceram. 8(1) (2019) 148-152.

[13] J. Gild, J. Braun, K. Kaufmann, E. Marin, T. Harrington, P. Hopkins, K. Vecchio, J. Luo, A high-entropy silicide: (Mo0.2Nb0.2Ta0.2Ti0.2W0.2)Si2, J. Materiomics 5(3) (2019) 337-343.

[14] X. Chen, Y. Wu, High-entropy transparent fluoride laser ceramics, Journal of the American Ceramic Society 103(2) (2020) 750-756.

[15] A.J. Wright, Q. Wang, C. Huang, A. Nieto, R. Chen, J. Luo, From high-entropy ceramics to compositionally-complex ceramics: A case study of fluorite oxides, J. Eur. Ceram. Soc. 40(5) (2020) 2120-2129.

[16] A.J. Wright, J. Luo, A Step Forward from High-Entropy Ceramics to Compositionally-Complex Ceramics: A New Perspective, Journal Materials Science 1000 (2020) preprint arXiv:2002.05251.

[17] J. Gild, K. Kaufmann, K. Vecchio, J. Luo, Reactive flash spark plasma sintering of high-entropy ultrahigh temperature ceramics, Scr. Mater. 170 (2019) 106-110.

[18] Y. Zhang, Z.-B. Jiang, S.-K. Sun, W.-M. Guo, Q.-S. Chen, J.-X. Qiu, K. Plucknett, H.-T. Lin, Microstructure and mechanical properties of high-entropy borides derived from boro/carbothermal reduction, J. Eur. Ceram. Soc. 39(13) (2019) 3920-3924.

[19] J. Gu, J. Zou, S.K. Sun, H. Wang, W. Weimin, Dense and pure high-entropy metal diboride ceramics sintered from self-synthesized powders via boro/carbothermal reduction approach, Sci. China Chem. (2019) 1898-1909.

[20] G. Tallarita, R. Licheri, S. Garroni, R. Orrù, G. Cao, Novel processing route for the fabrication of bulk high-entropy metal diborides, Scr. Mater. 158 (2019) 100-104.

[21] J. Gild, A. Wright, K. Quiambao-Tomko, M. Qin, J.A. Tomko, M. Shafkat bin Hoque, J.L. Braun, B. Bloomfield, D. Martinez, T. Harrington, K. Vecchio, P.E. Hopkins, J. Luo, Thermal conductivity and hardness of three single-phase high-entropy metal diborides fabricated by borocarbothermal reduction and spark plasma sintering, Ceram. Int. (2019).

[22] G. Tallarita, R. Licheri, S. Garroni, S. Barbarossa, R. Orrù, G. Cao, High-entropy transition metal diborides by reactive and non-reactive spark plasma sintering: a comparative investigation, J. Eur. Ceram. Soc. (2019).

[23] X.-F. Wei, J.-X. Liu, F. Li, Y. Qin, Y.-C. Liang, G.-J. Zhang, High entropy carbide ceramics from different starting materials, J. Eur. Ceram. Soc. 39(10) (2019) 2989-2994.

[24] X.-Q. Chen, C.L. Fu, M. Krčmar, G.S. Painter, Electronic and Structural Origin of Ultraincompressibility of 5d Transition-Metal Diborides MB2 (M = W, Re, Os), Phys. Rev. Lett. 100(19) (2008) 196403.

[25] E. Zhao, J. Meng, Y. Ma, Z. Wu, Phase stability and mechanical properties of tungsten borides from first principles calculations, Phys. Chem. Chem. Phys. 12(40) (2010) 13158-13165.

[26] C. Jiang, Z. Pei, Y. Liu, J. Xiao, J. Gong, C. Sun, Preparation and characterization of superhard AlB2-type WB2 nanocomposite coatings, Phys. Status Solidi A 210(6) (2013) 1221-1227.

[27] Y.M. Liu, R.Q. Han, F. Liu, Z.L. Pei, C. Sun, Sputtering gas pressure and target power dependence on the microstructure and properties of DC-magnetron sputtered AlB2-type WB2 films, J. Alloys Compd. 703 (2017) 188-197.

[28] Y.-M. Liu, T. Li, F. Liu, Z.-L. Pei, Thermal stability of WB2 and W–B–N films deposited by magnetron sputtering, Acta Metall. Sin.-Engl. 32(1) (2019) 136-144.

[29] H.P. Woods, F.E. Wawner, B.G. Fox, Tungsten diboride: preparation and structure, Science 151(3706) (1966) 75-75.

[30] J. Che, Y. Long, X. Zheng, H.-T. Lin, K. Plucknett, Mechanical alloying assisted spark plasma sintering of tungsten diboride ceramics, Mater. Chem. Phys. 237 (2019) 121848.

[31] M. Shibuya, M. Kawata, M. Ohyanagi, Z.A. Munir, Titanium diboride–tungsten diboride solid solutions formed by induction-field-activated combustion synthesis, J. Am. Ceram. Soc. 86(4) (2003) 706-10.

[32] M. Shibuya, Y. Yamamoto, M. Ohyanagi, Simultaneous densification and phase decomposition of TiB2–WB2 solid solutions activated by cobalt boride addition, J. Eur. Ceram. Soc. 27(1) (2007) 307-312.

[33] H. Kaga, E.M. Heian, Z.A. Munir, C. Schmalzried, R. Telle, Synthesis of hard materials by field activation: the synthesis of solid solutions and composites in the TiB2–WB2–CrB2 system, J. Am. Ceram. Soc. 84(12) (2001) 2764-2770.

[34] Y. Zhou, H. Xiang, Z. Feng, Z. Li, General trends in electronic structure, stability, chemical bonding and mechanical properties of ultrahigh temperature ceramics TMB2 (TM = transition metal), J. Mater. Sci. Technol. 31(3) (2015) 285-294.

[35] Y. Wang, M. Yao, Z. Hu, H. Li, J.-H. Ouyang, L. Chen, S. Huo, Y. Zhou, Microstructure and mechanical properties of TiB2-40 wt% TiC composites: effects of adding a low-temperature hold prior to sintering at high temperatures, Ceram. Int. 44(18) (2018) 23297-23300.

[36] W.G. Fahrenholtz, G.E. Hilmas, S.C. Zhang, S. Zhu, Pressureless Sintering of Zirconium Diboride: Particle Size and Additive Effects, J. Am. Ceram. Soc. 91(5) (2008) 1398-1404.

[37] E. Rudy, S. Windisch, Ternary Phase Equilibria In Transition Metal-Boron-Carbon-Silicon Systems. Part I. Related Binary Systems. Volume 7-9, Air Force Materials Laboratory, Wright-Patterson Air Force Base, OH, 1966.

[38] A. Jain, S.P. Ong, G. Hautier, W. Chen, W.D. Richards, S. Dacek, S. Cholia, D. Gunter, D. Skinner, G. Ceder, K.A. Persson, Commentary: the Materials Project: a materials genome approach to accelerating materials innovation, APL Mater. 1(1) (2013) 011002.

[39] J. Schmidt, M. Boehling, U. Burkhardt, Y. Grin, Preparation of titanium diboride TiB2 by spark plasma sintering at slow heating rate, Sci. Technol. Adv. Mat. 8(5) (2007) 376-382.

[40] S. Ran, L. Zhang, O. Van der Biest, J. Vleugels, Pulsed electric current, in situ synthesis and sintering of textured TiB2 ceramics, J. Eur. Ceram. Soc. 30(4) (2010) 1043-1047.

[41] F.K. Lotgering, Topotactical reactions with ferrimagnetic oxides having hexagonal crystal structures—I, J. Inorg. Nucl. Chem. 9(2) (1959) 113-123.

[42] W.G. Fahrenholtz, G.E. Hilmas, I.G. Talmy, J.A. Zaykoski, Refractory Diborides of Zirconium and Hafnium, J. Am. Ceram. Soc. 90(5) (2007) 1347-1364.

[43] L. Silvestroni, D. Sciti, Effects of MoSi2 additions on the properties of Hf– and Zr–B2 composites produced by pressureless sintering, Scr. Mater. 57(2) (2007) 165-168.

[44] L. Silvestroni, S. Guicciardi, C. Melandri, D. Sciti, TaB2-based ceramics: Microstructure, mechanical properties and oxidation resistance, J. Eur. Ceram. Soc. 32(1) (2012) 97-105.

[45] S.C. Zhang, G.E. Hilmas, W.G. Fahrenholtz, Oxidation of Zirconium Diboride with Tungsten Carbide Additions, J. Am. Ceram. Soc. 94(4) (2011) 1198-1205.

[46] S.C. Zhang, G.E. Hilmas, W.G. Fahrenholtz, Improved Oxidation Resistance of Zirconium Diboride by Tungsten Carbide Additions, J. Am. Ceram. Soc. 91(11) (2008) 3530-3535.

[47] P. Vajeeston, P. Ravindran, C. Ravi, R. Asokamani, Electronic structure, bonding, and ground-state properties of AlB2-type transition-metal diborides, Phys. Rev. B 63(4) (2001) 045115.